\pdfoutput=1
\documentclass[preprint,aps,pre]{revtex4-1}
\usepackage[utf8]{inputenc}
\usepackage{epsfig}
\usepackage{subfigure}
\usepackage{latexsym}
\usepackage{amsmath}
\usepackage{psfrag}
\usepackage{fullpage}
\usepackage{graphicx}
\usepackage{amssymb}
\usepackage{setspace}

\DeclareMathOperator{\rank}{rank}

\begin{document}
 \title{Phase transition in the controllability of temporal networks}

\author{M\'arton P\'osfai}
\affiliation{Department of Physics of Complex Systems, E\"otv\"os University, P\'azm\'any P\'eter s\'et\'any 1/A, H-1117 Budapest, Hungary}
\affiliation{ Institut f{\"u}r Theoretische Physik, Technische Universit\"at 
Berlin, Hardenbergstra\ss{}e 36, 10623 Berlin, Germany}

\author{Philipp H\"ovel}
\affiliation{Institut f{\"u}r Theoretische Physik, Technische Universit\"at 
Berlin, Hardenbergstra\ss{}e 36, 10623 Berlin, Germany}
\affiliation{Bernstein Center for Computational Neuroscience, 
Humboldt-Universit{\"a}t zu Berlin, Philippstra{\ss}e 13, 10115 Berlin, Germany}

\date{\today}

\begin{abstract}
The control of complex systems is an ongoing challenge of complexity research. Recent advances using concepts of structural control deduce a wide range of control related properties from the network representation of complex systems. Here, we examine the controllability of complex systems for which the timescale of the dynamics we control and the timescale of changes in the network are comparable. We provide both analytical and computational tools to study controllability based on temporal network characteristics. We apply these results to investigate the controllable subnetwork using a single input, present analytical results for a generic class of temporal network models, and preform measurements using data collected from a real system. Depending upon the density of the interactions compared to the timescale of the dynamics, we witness a phase transition describing the sudden emergence of a giant controllable subspace spanning a finite fraction of the network. We also study the role of temporal patterns 
and network topology in real data making use of various randomization procedures, finding that the overall activity and the degree distribution of the underlying network are the main features influencing controllability.
\end{abstract}

\maketitle

\section{Introduction}

Complex systems consist of many interacting elements, and the web of these interactions are best described by a complex network. Therefore, studying the structure of such networks and exploring the consequences of their properties is essential to understand complexity. In the last two decades, significant amount of research has been devoted to this problem~\cite{ALB02a, NEW03, BAR08c, COH10, FOR10}, however, only limited progress has been made in describing how the network structure of the system influences our ability to control it~\cite{WAN02a, LOM07, YU09, FIE13}. Recent work by Liu et al. spurred interest in network control~\cite{LIU11}. They found that, if the system can be represented by a directed weighted network, assuming linear dynamics and invoking the framework of structured systems~\cite{LIN74, HOS80}, it is possible to study control related questions by only using information about the underlying network. This enabled the research community to apply the full arsenal of network science to the problem, uncovering various nontrivial phenomena emerging from the complexity of the structure of the system~\cite{NEP12, POS13, JIA13, SUN13, YUA13}.

\begin{figure}
	\centering
	\includegraphics[scale=.5]{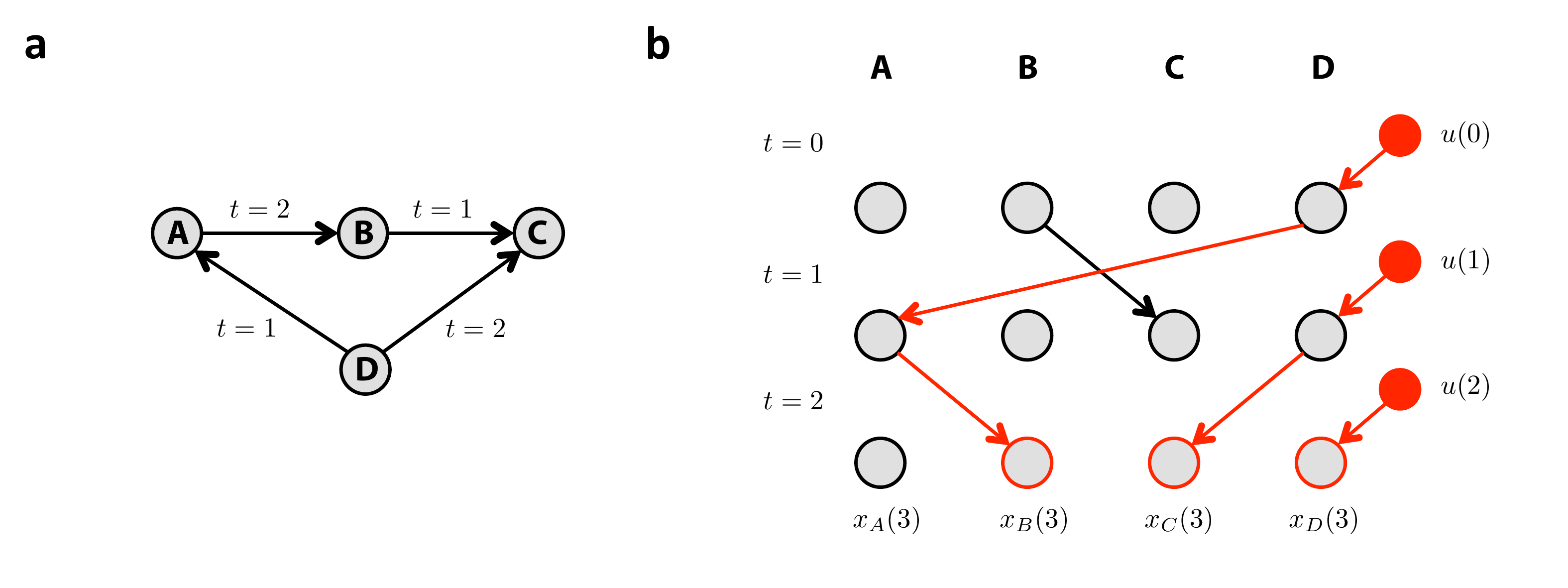}
	\caption{{\bf Controlling a simple temporal network.} (a) In the aggregated network, information can spread from node A to C, however, it is forbidden if the temporal order of the interactions is taken into account. As a consequence, we cannot control C by imposing a control signal on A. (b) We visualize the dynamics represented by the small temporal network by creating a copy of each node for each time step. The state of the node in the layer $t+1$ is determined by its neighbors in layer $t$. We aim to control the system at $t=3$ in $\Delta t= 3$ time steps. We use D as an input node, meaning that we can intervene at D in layers $t=1,2,3$. According to the independent path theorem, we can control the nodes in layer $t=3$ that can be connected to intervention points via independent paths, therefore we control nodes B, C and D.}
	\label{fig:example}
\end{figure}

In this paper, we extend structural controllability to systems for which the timescale of the dynamics and the timescale of changes in the network topology are comparable~\cite{HOL12}. In particular, it is necessary to take into account temporal information of the connections when the interaction events are not evenly distributed over time, but have nontrivial temporal correlations~\cite{BAR05,MAL08a}. Such systems include communication, trade, or transportation networks~\cite{VAZ07, IRI09, JO12, PAN11, KON12a}. Furthermore, the temporal sequence, of interactions governs spreading processes~\cite{VAZ07, IRI09}. Consider an example, a small communication network of three individuals $A$, $B$ and $C$ (Fig.~\ref{fig:example}a). Assume that $B$ sends an email to $C$ at time $t=1$, and $A$ sends an email to $B$ at time $t=2$. Neglecting the temporal sequence, we find that information may spread from $A$ to $C$. However, taking the order of the messages into account, this is obviously not possible, which has a clear consequence for control: we cannot influence $C$ using $A$. Therefore, one must include the temporal aspect of the interactions, when studying the controllability of networks with time-varying topologies.

\section{Structural controllability of temporal networks}

We study directed temporal networks $\mathcal{T}$, which are composed of a set nodes $V=\{v_1, v_2, \ldots, v_N\}$ and a set temporal links $E=\{e_1, e_2, \ldots, e_L\}$. Each temporal link $e=(v_i,v_j,t)\in E$ consists of an ordered node pair and a time stamp, representing that the node $v_i$ interacts with node $v_j$ at time $t$. Furthermore, we assume that the links are weighted, although the weight does not have to be known.

We consider discrete time-varying linear dynamics~\cite{KWA72}
\begin{equation}
\label{eq:model}
 x(t+1)={\bf A}(t) x(t) + {\bf B}(t)u(t),
\end{equation}
where the vector $x(t)\in \mathbb{R}^N$ represents the state variables, $x_i(t)$ corresponding to the state of node $v_i$ at time $t$. The first term describes the internal dynamics of the system, the  matrix ${\bf A}(t)\in \mathbb{R}^{N\times N}$ is the transpose of the weighted adjacency matrix at time $t$. The second term describes the control applied to the system: if we impose an outside signal on node $v_i$ at time $t$ changing the state of the node at time $t+1$, we say that we intervene at node $v_i$, and we call the $(v_i,t+1)$ pair an intervention point. The vector $u(t)\in \mathbb{R}^{N_\text I(t)}$ is a list of interventions, where $N_\text I(t)$ is the number of interventions at time $t$. The nonzero elements of matrix ${\bf B}(t)\in \mathbb{R}^{N\times N_\text I(t)}$ identify the intervention points.

Extending the standard definition of structural controllability to time-varying systems~\cite{LIN74, HAR13a}, we call a subset of nodes $C\subseteq V$ a structural controllable subspace at target time $t$ in $\Delta t$ time steps, if there exists a pair of ${\bf A}^*(t)$ and ${\bf B}^*(t)$ that has the same structure as ${\bf A}(t)$ and ${\bf B}(t)$, such that the state of all nodes $v\in C$ can be driven from any initial state to any final state at time $t$, in at most $\Delta t$ time steps by appropriately choosing $u(t)$. By same structure we mean that the zero entries in ${\bf A}(t)$ and ${\bf A}^*(t)$ are in the same places and only the value of the nonzero elements can be different, i.e. the links connect the same nodes in the corresponding network, only the weights can be different. It is worth noting that explicitly including $\Delta t$ in the definition allows us to study the time necessary to achieve control, an aspect that has not been explored yet.

The power of the structural controllability approach arises from the fact that it does not require detailed information about the strength of the interactions, allowing us to characterize controllability by considering network properties only. Yet, structural controllability is a general property, in the sense that if a system is structural controllable, it is controllable for almost all weight configurations~\cite{LIN74, HOS80}. 

We prove the independent path theorem in the Supplementary Information. The theorem states that $C$ is a structurally controllable subspace, if all nodes $v\in C$ at time $t$ are connected to intervention points through independent time-respecting paths of length of at most $\Delta t-1$. A time-respecting path is a sequence of adjacent temporal links such that subsequent links in the path are active in subsequent time steps, e.g. the link $(v_i,v_j,t)$ may be followed by $(v_j,v_k,t+1)$. Two paths are independent if they do not pass the same node at the same time. For a small example see Fig.~\ref{fig:example}.

The independent path theorem allows us to formulate control related questions, here we focus on the problem of identifying the maximum controllable subspace $N_\text C (v,t,\Delta t)$ using a single input node $v$, i.e. we allow interventions at points $(v,s)$ for any $t-\Delta t < s \leq t$. To determine $N_\text C (v,t,\Delta t)$, we have to find the maximum number of independent paths starting from possible intervention points and ending at time $t$, i.e. ending at points $(w,t)$ for any $w\in V$. Identifying the independent paths can be done efficiently using the Ford-Fulkerson algorithm as explained in the Supplementary Information. We characterize the overall controllability of a temporal network by the average maximum controllable subspace
\begin{equation}
 N_\text C (t,\Delta t) = \frac{1}{N} \sum_{v\in V} N_\text C (v,t,\Delta t).
\end{equation}

\section{Analytical results for model networks}

We provide an analytical solution for a simple class of model networks to gain insight on the effect of the degree distribution and the choice of $\Delta t$. To create the network, we generate an uncorrelated static network for each time step with prescribed in- and out-degree distributions $p_\text{in}(k)$ and $p_\text{out}(k)$, respectively. Each time step is generated independently, only the degree distributions are kept the same. Therefore, consecutive time steps are completely uncorrelated. Since all time steps are statistically equivalent, the maximum controllable subspace does not depend on $t$, i.e. $N_\text C (t,\Delta t)\equiv N_\text C (\Delta t)$. We study networks with Poisson (Erd{\H o}s-R{\'e}nyi networks)~\cite{ERD59} and scale-free degree distributions~\cite{BAR99a}, the latter meaning that $p_\text{in/out}(k)\sim k^{-\gamma_\text{in/out}}$.

Consider first the case of only one intervention point $(v,s)$. We can use this intervention to control one of the accessible nodes at later time steps, i.e. any node that can be reached via a path originating from $(v,s)$. The cluster of accessible nodes can be described as  generated by the Galton-Watson branching process~\cite{HAR63}: node $v$ at time $s$ has $k_\text{out}$ offspring, $k_\text{out}$ is drawn from the distribution $p_\text{out}(k)$, each of these offspring have $k_\text{out}$ out-neighbors also drawn independently from  $p_\text{out}(k)$, and so forth. The Galton-Watson process undergoes a phase transition depending on the average degree: in the subcritical phase $\langle k \rangle<1$, it will terminate in finite steps, reaching only a finite number of nodes; in the supercritical phase $\langle k \rangle>1$, and the branching process may continue forever, spanning a finite fraction of the network. We will show in the following that the existence of infinite long paths fundamentally changes the controllability of the system.

\begin{figure}
	\centering
	\includegraphics[scale=.5]{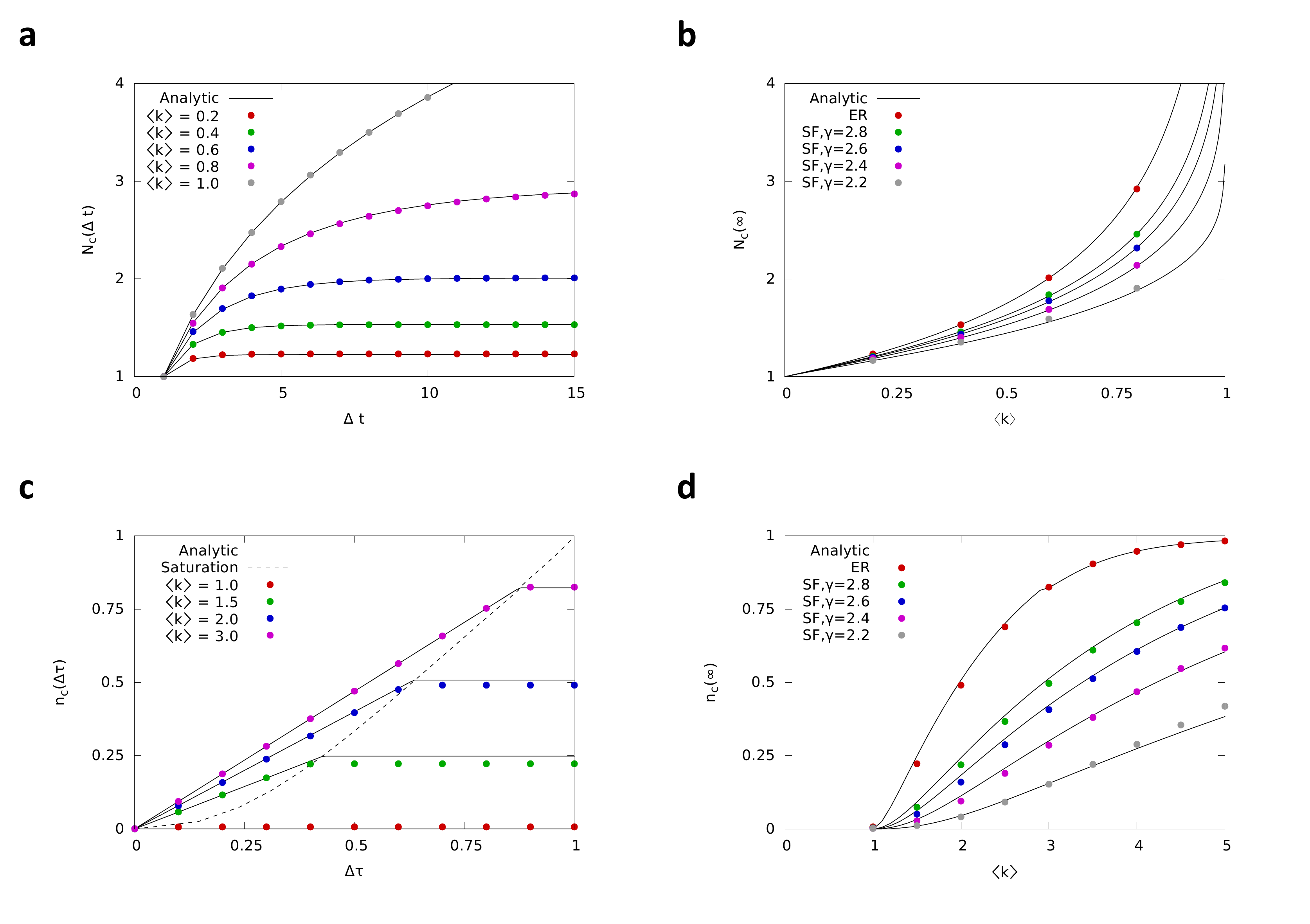}
	\caption{{\bf Temporal controllability of model networks.} (a) The average maximum controllable subspace $N_\text C(\Delta t)$ for Poisson distributed networks in the subcritical phase. The $N_\text C(\Delta t)$ reaches its stationary value exponentially fast, and increasing the average degree increases both $N_\text C(\infty)$ and the time necessary for saturation. In the critical point $\langle k\rangle=1$, $N_\text C(\Delta t)$ does not saturate for finite $\Delta t$, but increases logarithmically. (b) We show the stationary value of the average maximum controllable subspace $N_\text C(\infty)$ in function of the average degree for Poisson and scale-free networks in the subcritical phase. The existence of hubs make control increasingly difficult, the difference between Poisson and scale-free networks is the most prominent in the critical point, where $N_\text C(\infty)$ diverges for Poisson distributed networks, but remains finite for scale-free networks. (c) In the supercritical phase, we can control a 
finite fraction of the network. We show $n_\text C(\Delta\tau)= N_\text C(\Delta\tau N)/N$ for Poisson networks. For small $\Delta\tau$, $n_\text C(\Delta\tau)$ increases linearly, saturates at a characteristic $\Delta\tau^*$ (dashed line), and remains constant for larger $\Delta\tau$. (d) We plot the $n_\text C(\infty)$ for Poisson and scale-free networks, and again find that heterogeneity of the degree distribution lowers $n_\text C(\infty)$. The dots are results of simulations for networks of size $N=1,000$, each data point is an average of $10,000$ input node measurements in the subcritical phase, and $1,000$ in the supercritical phase. The continuous line shows the analytic solution, for the scale-free networks we have taken the finite size effect into account (SI Sec.~II).}
	\label{fig:model}
\end{figure}

In the subcritical regime, we find that
\begin{equation}
\label{eq:subcrit_solution}
 N_\text C (\Delta t) = 1+\sum_{t=0}^{\Delta t-2}1-P(t),
\end{equation}
where $P(d)$ is the cumulative distribution function of the maximum path length originating from an intervention point. $P(d)$ is determined using a self-consistent recursive formula, and only depends on the out-degree distribution $p_\text{out}(k)$. For long control times $\Delta t \rightarrow \infty$, equation (\ref{eq:subcrit_solution}) predicts $N_\text C (\infty) = 1 + \langle d \rangle$, which is simply one larger than the average maximum path length. For the same average degree, $\langle d \rangle_\text{ER}$ of Poisson distributed networks is always larger than $\langle d \rangle_\text{SF}$ in scale-free networks. Furthermore, in the scale-free case $\langle d \rangle_\text{SF}$ decreases as the degree exponent $\gamma_\text{out}$ decreases, evincing that the presence of hubs makes control increasingly difficult. The infinite solution is approached exponentially fast for both Poisson and scale-free networks with characteristic time $\Delta t^*=1/\lvert\log\langle k \rangle\rvert$. This means that only few time 
steps are needed for maximum controllability. (Fig.~\ref{fig:model}a,b)

In the transition point $\langle k \rangle=1$,  $\langle d \rangle_\text{ER}$ diverges, and $N_\text C^\text{ER} (\Delta t)\sim \log \Delta t$ for large $\Delta t$. For scale-free networks $\langle d \rangle_\text{SF}$ remains finite, but
the asymptotic solution is reached slower $N_\text C^\text{SF}(\infty)-N_\text C^\text{SF}(\Delta t)\sim \Delta t^{-(3-\gamma_\text{out})/(\gamma_\text{out}-2)}$. Finite size, however, can obscure the difference between the two network classes by introducing a cutoff in the degree distribution.

Above the critical point $\langle k \rangle>1$, the maximum path length is no longer a limitation due to the formation of a giant component. Consequently, choosing large $\Delta t$, i.e. $\Delta t= \Delta\tau N$ ($\Delta\tau>0$), we can control a finite fraction of the network $n_\text C(\Delta\tau)=N_\text C(\Delta\tau N) / N$.

For small $\Delta\tau$, any infinite path starting from an intervention point can be used for control. Therefore,
\begin{equation}
\label{eq:supercrit_solution}
 n_\text C (\Delta\tau) = \Delta\tau S_\text{out},
\end{equation}
where $S_\text{out}$ is the probability that an intervention point is a root of an infinite path, which is determined by a self-consistent equation. Similarly to the subcritical regime, the solution only depends upon $p_\text{out}(k)$.

Examining equation (\ref{eq:supercrit_solution}), one might think that by allowing sufficiently large $\Delta\tau$, we can control the entire network. However, above a characteristic $\Delta\tau^*$, a new limitation arises, and $n_\text C (\Delta\tau)$ saturates (Fig.~\ref{fig:model}c). The number of controlled nodes will be equal to the maximum number of independent infinite paths, i.e. infinite paths that do not pass the same node at the same time step. We analytically approximate $n_\text C (\infty)$ using the framework developed to study core percolation and maximum matching in the Supplementary Information~\cite{LIU12, ZDE06}. We find that $n_\text C (\infty)$ depends on both $p_\text{in}(k)$ and $p_\text{out}(k)$, and it is symmetric to swapping the two distributions: it does not matter which direction we follow the paths, the number if independent paths remains the same. 

Comparing the Poisson and scale-free distributions, we find that the Poisson distributed networks are easier to control both below and above the saturation point $\Delta\tau^*$, in line with our observation in the subcritical regime (Fig.~\ref{fig:model}b,d).

\section{Temporal controllability of a real system}

Digital traces of communication make it possible to apply the developed tools to explore the controllability of real systems. Here,  we study a temporal network representing the email communication of a mid-size company~\cite{MIC11}. The dataset contains the sender, the recipient, and the time each email has been sent. All together there are 82,927 emails between 167 employees covering a 9 month period. The necessary temporal resolution of the network depends on the timescale of the dynamical process we aim to control. To highlight different features of the dataset we use two different temporal resolutions with one hour and one day time steps. The first corresponds to a short term control scenario, influencing the dynamics within a workday, while the second case assumes a slower change, spanning the whole available period. The dataset features strong daily and weekly patterns: the bulk of the email traffic happens during a 9 hour period of the workdays. Therefore, for the short term control case we average the results 
for workdays only, and for the long term case we remove the weekends and holidays.

\begin{figure}
	\centering
	\includegraphics[scale=.4]{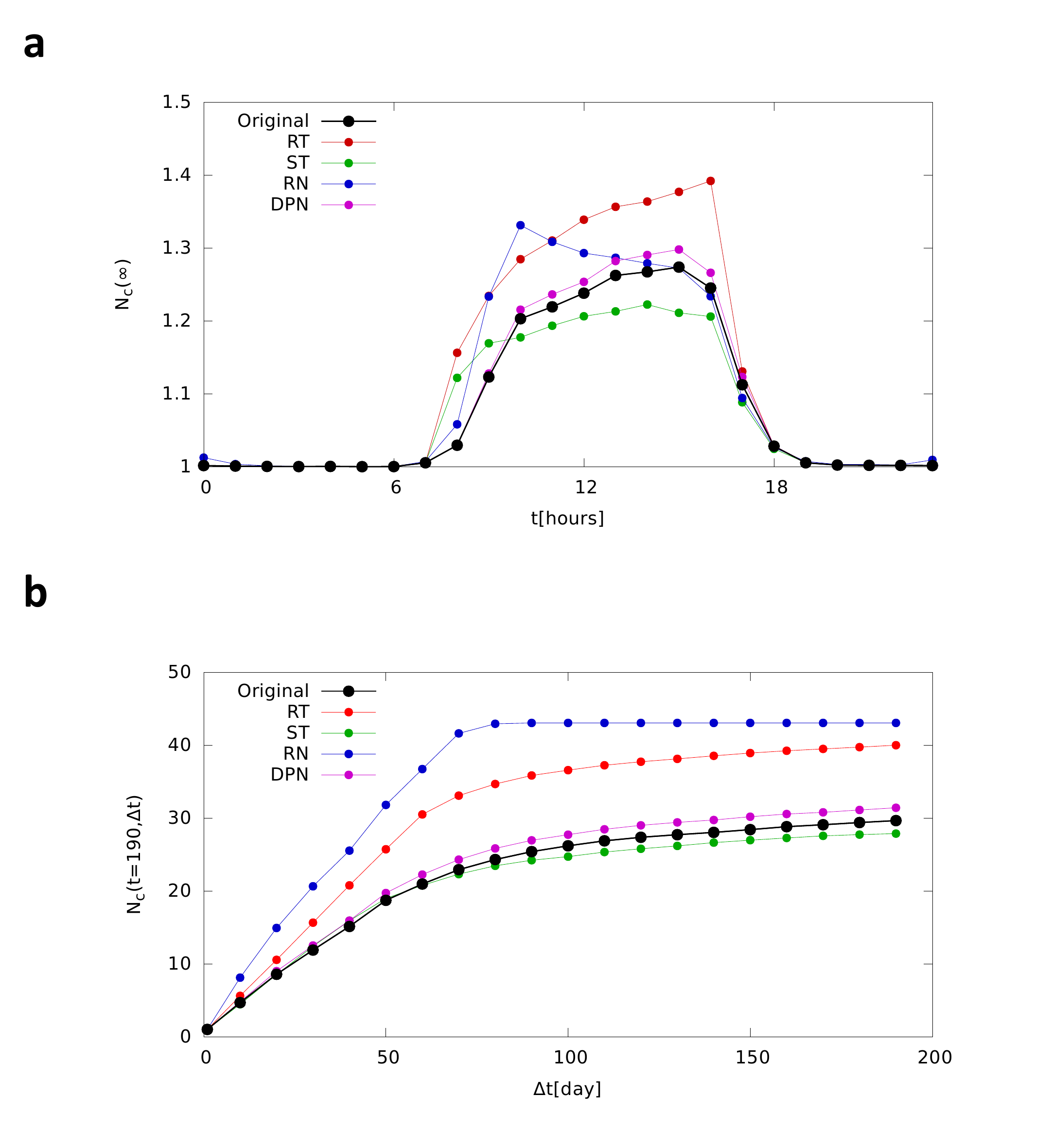}
	\label{fig:real}
	\caption{{\bf Controllability of a real system} (a) In the first scenario, we aim to control the company within a workday, thus we set the temporal resolution to 1 hour. The average degree of the network within a time step is $\langle k \rangle_\text h \approx 0.23<1$, that is the system is in the subcritical phase. We show $N_\text C (t,\Delta t= \infty)$, i.e. the average maximum controllable subspace in function of the target time $t$ in the $\Delta t\rightarrow\infty$ limit. The data points are the average of 189 workdays covered by the dataset. The bulk of the email traffic happens during the working hours. Therefore, we restrict ourselves to the regular office hours spanning the period $9\leq t\leq17$ when randomizing the time of the links. Completely removing temporal patterns by assigning random times to the links (RT), and completely removing the network structure by randomly placing the links within a time step (RN) both increase $N_\text C(t,\infty)$. Shuffling the time steps (ST) only 
slightly increases, and randomizing the links while keeping the degree sequence (DPN) slightly decreases $N_\text C(t,\infty)$. This shows that the controllability is mainly determined by the degree distribution and the overall activity pattern, correlations have smaller impact. (b) In the second scenario, we aim to control the system on a longer time scale, we chose 1 workday as a unit of time. In this case the average degree within a time step is $\langle k \rangle_\text d \approx 1.76>1$, predicting that the system is above the critical point, and that the characteristic time to control the system will be in the order of the system size. Therefore, we cannot make multiple independent measurements, and we will focus on control at the end of the last workday at $t=190$. We show the $N_\text C(t=190, \Delta t)$ in function of $\Delta t$. We observe linear growth for small $\Delta t$ and saturation for large $\Delta t$, although the saturation does not happen completely in the available time period. 
Randomizations yield similar conclusion as in the short term control scenario.}
\end{figure}

The average degree of the network in one time step depends on the time resolution. For the one hour time steps we find $\langle k \rangle_\text h \approx 0.23$, predicting that system is in the subcritical phase. Indeed, we find that $N_\text C(t,\Delta t)$ remains of the order of few nodes, and it saturates in just a few steps in accordance with our findings for model networks (Fig.~\ref{fig:real}a). For the one day time step, we obtain $\langle k \rangle_\text d \approx 1.76$, putting the system in the supercritical regime. We find that in the beginning $N_\text C(t, \Delta t)$ increases approximately linearly with $\Delta t$, and for larger $\Delta t$ it seems to saturate, although slower than in the case of model networks (Fig.~\ref{fig:real}b).

Next, we use various randomization processes to separate the effects of temporal patterns and the underlying network. We find that fluctuations in the average degree of the time steps decreases $N_\text C(t,\Delta t)$: a drop in the average degree acts as a bottleneck, letting through fewer independent paths. Indeed, removing the fluctuations by assigning random times to the links, significantly increases $N_\text C(t,\Delta t)$ (RT curve in Fig.~\ref{fig:real}). Next, we shuffle the time steps, meaning that we keep the overall fluctuations in the average degree, but we eliminate the correlations between subsequent time steps. We find that $N_\text C(t,\Delta t)$ slightly decreases, suggesting that temporal correlations enhance the number of available paths, such as casual chain of events (ST curve). To investigate the effect of the underlying network, we keep the temporal information, and we only randomize the network within a time step. First, we completely mix the connections, thereby transforming the 
degree distribution to a Poisson distribution with the same average degree. The controllability of the resulting network dramatically increases, showing that the existence of hubs makes control difficult (RN curve). In the next randomization, we keep the degree of each node in each time step, but we eliminate all other correlations by cutting all links and randomly rewiring them. We find that the controllability of such networks is very close to the original, meaning that the degree sequence of the nodes is the main factor in determining controllability, correlations are only secondary (DPN curve). 

\section{Conclusions}

Both structural controllability and temporal networks proved to be a useful tool in understanding complex systems, generating a high amount of research in their respective fields. Here, we have established the connection  between the two, opening an array of new questions. We explored how the overall activity and the degree distribution of the underlying network influence controllability. Further questions, such as the minimum set of input nodes necessary for complete control or the role of individual nodes are yet to be explored.

\begin{acknowledgments}
This project was supported by German Academic Exchange Service (DAAD) via a scholarship granted to MP, and PH acknowledges support by BMBF (grant no. 01Q1001B) in the framework of BCCN Berlin.
\end{acknowledgments}

\section{Supplementary Information}

\subsection{Structural controllability of temporal networks}

\subsubsection{Temporal networks}

 A directed temporal network $\mathcal{T}$ is defined as a set nodes $V=\{v_1, v_2, \ldots, v_N\}$ and a set temporal links $E=\{e_1, e_2, \ldots, e_L\}$~\cite{HOL12}. Each temporal link $(v_j,v_k,t)\in E$ consists of an ordered node pair and a time stamp, representing that there is a link pointing from node $v_j$ to node $v_k$ at time $t$. We measure the time in discrete steps $t=0,1,2\dots$, the choice of the unit may depend on the resolution of the available dataset or modelling purposes. Furthermore, we assume that each link $e\in E$ has a weight $w_e\in \mathbb{R}$ associated to it, although the weight is not necessarily known.
 
\begin{figure}
	\centering
	\includegraphics[scale=.5]{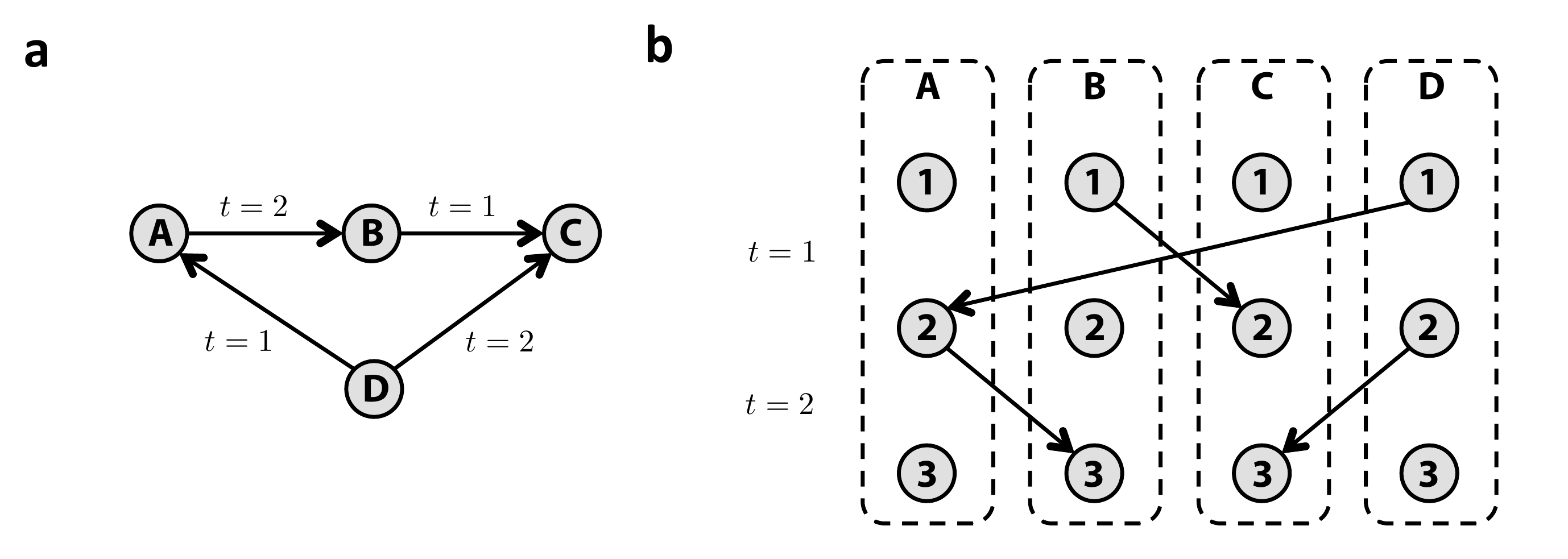}
	\caption{{\bf Layered network example.} (a) A small temporal network of 4 nodes. In the static representation of the network, information can spread from node $A$ to node $C$, however, due to the temporal sequence of the interactions, this is not possible. There is a time respecting path from $D$ to $B$ (consisting of links $(D,A,1)$ and $(A,B,2)$), and from $D$ to $C$ (consisting of link $(D,C,2)$). The two paths do not pass the same node at the same time, therefore they are independent. (b) The layered network representation $\mathcal L(t_0=0,t_1=3)$. We make a copy $\hat v_{(i,t)}$ of each node $v_i$ for each time step $t\in(t_0,t_1]$. We connect the two nodes $\hat v_{(i,t)}$ and $\hat v_{(j,t+1)}$ if there exist a temporal link connecting nodes $v_i$ and $v_j$ at time $t$. There are altogether $\lvert \hat V \rvert = \lvert V \rvert(t_1-t_0)=12$ nodes in $\mathcal L(0,3)$.}
	\label{fig:layered_def}
\end{figure}

If there exists a link $(v_i,v_j,t)\in E$, then $v_i$ is the in-neighbor of $v_j$, and $v_j$ is the out-neighbor of $v_i$. The temporal links $(v_i,v_j,t)$ and $(v_k,v_l, s)$ are consecutive, if $j=k$ and $t+1=s$. A temporal path $P$ connecting node $v_i$ and $v_j$ from $t_0$ to $t_1$ is a sequence of consecutive temporal links such that the first link originates from node $v_i$ at time $t_0$, and the last link in the sequence points at node $v_j$ at time $t_1$. The path consist of $\Delta t = t_1 - t_0 $ consecutive links and $\Delta t + 1$ nodes. A node by itself is a path of length $0$. Two paths are independent if they do not pass through the same node at the same time. For a small example see Fig.~\ref{fig:layered_def}a.

\subsubsection{Layered network representation}
 
It will be useful to represent the temporal network defined above as a layered network $\mathcal L(t_0,t_1)$ consisting of a set of nodes $\hat V$ and a set of static links $\hat E$. We make a copy $\hat v_{(i,t)}\in \hat V$ of each node $v_i \in V$ for each time step $t\in(t_0,t_1]$. We connect the two nodes $\hat v_{(i,t)}$ and $\hat v_{(j,t+1)}$ if there exist a temporal link $(v_i,v_j,t)$ (See Fig.~\ref{fig:layered_def}b). Therefore, the layered representation is a static directed acyclic network with $\lvert \hat V \rvert=\lvert V \rvert(t_1-t_0)$ nodes.
 
As a consequence, temporal paths appear as static paths in the layered representation, and independent temporal paths are simply node-disjunct paths.
 
\subsubsection{Dynamics}

We study discrete time linear dynamics~\cite{KWA72}
\begin{equation}
\label{eq:model_si}
 x(t+1)={\bf A}(t) x(t) + {\bf B}(t)u(t),
\end{equation}
where the vector $x(t)\in \mathbb{R}^N$ represents the state variables, $x_i(t)$ corresponding to the state of node $v_i$ at time $t$. The  matrix ${\bf A}(t)\in \mathbb{R}^{N\times N}$ provides information about the interactions at time $t$: if there exists a link $(v_i,v_j,t)$ with associated weight $w$, the matrix element $a_{ji}(t)=w$, otherwise $a_{ji}(t)=0$. The vector $u(t)\in \mathbb{R}^{N_\text I(t)}$ is the control signal, we call each element of $u(t)$ an intervention, and $N_\text I(t)$ is the number of interventions at time $t$. The matrix ${\bf B}(t)\in \mathbb{R}^{N\times N_\text I(t)}$ tells us, at which node we intervene: $b_{ij}(t)\neq 0$ means that we shift $x_i(t+1)$ by $b_{ij}(t)u_j$. We define the total number of interventions $N_\text I=\sum_t N_\text I (t)$. If we intervene at a node at any time, the node is referred to as an input.

{\it Note 1:} The state $x_i(t+1)$ of node $v_i$ is completely determined by the state of its in-neighbors at time $t$. If we assume that $x_i(t+1)$ is not independent from $x_i(t)$, we have to add self-interactions (e.g. diagonal entries in ${\bf A}(t)$). Information about self-interactions is not always explicitly provided in network datasets. 

{\it Note 2:} Consider the case when there are no links pointing at $v_i$ at $t$, that is $a_{ij}(t)=0$ for all $j$. If we did not add self-interactions, then $x_i(t+1)=0$. In some cases, we might assume that if a node does not have incoming links, it retains its state. This can be taken into account by adding a self-interaction at time $t$ only if a node has no incoming links.

In this paper we focus on the general case, when self-interactions may or may not be present. The effect of self-loops is an open question left for future research.

\subsubsection{Controllability}

{\it Controllable:} We call the system $({\bf A}(t), {\bf B}(t))$ controllable at time $t_1$ in $\Delta t$ time steps, if the system can be driven to any final state $x(t_1)$ at time $t_1$ from any initial state $x(t_0)$ in at most $\Delta t=t_1-t_0$ time steps. 
Since the system is linear, we can transform $x(t)$ to set $x(t_0)=0$ without loss of generality.

By successively applying equation~(\ref{eq:model_si}), the state of the system at time $t_1$ is
\begin{multline}
\label{eq:state_at_t1}
 x(t_1) =  {\bf A}(t_1-1){\bf A}(t_1-2)\cdots {\bf A}(t_0+1){\bf B}(t_0) u + \ldots \\
 + {\bf A}(t_1-1){\bf B}(t_1-2)u + {\bf B}(t_1-1)u.
\end{multline}

We define the temporal contollability matrix~\cite{KWA72, HAR13a}
\begin{equation}
{\bf C}(t_0,t_1) = \left[ {\bf A}(t_1-1){\bf A}(t_1-2)\cdots {\bf A}(t_0+1){\bf B}(t_0);\ldots;{\bf A}(t_1-1){\bf B}(t_1-2); {\bf B}(t_1-1)\right],
\end{equation}
where $[{\bf X};{\bf Y}]$ is the concatenation of matrices ${\bf X}$ and ${\bf Y}$, therefore ${\bf C}(t_0,t_1)\in\mathbb{R}^{N\times N_\text I}$ with the total number of interventions $N_\text I=\sum_t N_\text I (t)$. Using this definition we simply get
\begin{equation}
 x(t_1) =  {\bf C}(t_0,t_1) u,
\end{equation}
where $u=[u(t_0)^T;u(t_0+1)^T;\ldots; u(t_1)^T ]^T\in\mathbb{R}^{N_\text I}$. It is now clear that the linear rank of ${\bf C}(t_0,t_1)$ is the number of variables that can be set independently by the proper choice of $u$, that is $({\bf A}(t), {\bf B}(t))$ is controllable if
\begin{equation}
 \rank {\bf C}(t-\Delta t,t) = N. 
\end{equation}

In most cases, however, the strength of the interactions, i.e. the link weights, are not known completely. Fortunately, a lot of information about the controllability of a system can be deduced only from the zero-nonzero structure of $({\bf A}(t), {\bf B}(t))$ , i.e. the existence or absence of links, using the structural controllability framework. We treat the nonzero elements of $({\bf A}(t), {\bf B}(t))$ as free parameters, and only keep the zero elements fixed. 

{\it Structurally controllable:} We call the system $({\bf A}(t), {\bf B}(t))$ structurally controllable at time $t$ in $\Delta t$ time steps, if we can set the free parameters of $({\bf A}(t), {\bf B}(t))$ such that the system is controllable in the original sense.

{\it Note:} If a system is structurally controllable, it is controllable for almost all weight configurations. And if it is not, it can be made controllable with an arbitrarily small perturbation of the weights~\cite{LIN74}.  

{\it Controllable subspace:} We call the subset of state variables $C\subseteq V$ a controllable subspace at time $t_1$ in $\Delta t$ time steps, if the state variables $x_i\in C$ can be driven to any final state at time $t_1$ from any initial state in at most $\Delta t=t_1-t_0$ time steps.

{\it Structurally controllable subspace:} We call the subset of state variables $C\subseteq V$ a structurally controllable subspace at time $t$ in $\Delta t$ steps, if we can set the free parameters of $({\bf A}(t), {\bf B}(t))$ such that $C$ is a controllable subspace in the original sense.

\subsubsection{Independent path theorem}
 

 {\bf Theorem:} $C\subseteq V$ is a controllable subspace of $\mathcal T(V,E)$ with dynamics $({\bf A}(t), {\bf B}(t))$ at time $t_1$ in $\Delta t$ time steps, iff there exists a $\lvert C \rvert$ independent paths starting from intervention points within $(t_1-\Delta t, t_1]$ and ending at nodes $v_i\in C$ at time $t_1$. 
 
 {\it Proof:}
 We reduce the time-dependent controllability problem to a larger time-independent problem
 \begin{equation}
  \hat x(\hat t+1) = \hat {\bf A}\hat x(\hat t) + \hat {\bf B} \hat u.
 \end{equation}
 We construct the linear time-independent system the following way: We create $\hat x \in \mathbb{R}^{N\cdot\Delta t}$ state vector, such that $\hat x_{(i,t)}(\hat t)$ corresponding to $x_i(t)$, $t\in(t_1-\Delta t, t_1]$. Note that we use the index pair $(i,t)$ to identify the elements of vector $\hat x$. We construct $\hat {\bf A}$ by setting $\hat a_{(i,t);(j,t+1)}$ to $a_{ij}(t)$, all other elements of $\hat {\bf A}$ are set to $0$. The input nodes in the time-independent system correspond to the intervention points of the time-dependent system, that is $\hat u=u$, and $\hat b_{(i,t);j}$ is $b_{i,j}(t)$. The network representation of the time-independent system is equivalent to the layered graph representation $\mathcal L(\hat V,\hat E,t_1-\Delta t, t_1)$ of the temporal network $\mathcal T(V,E)$.
 
 We can check by simple multiplication that $\hat x_{(i,t_1)}(\hat t = \Delta t)=x_i(t_1)$ for all $i$. Therefore for every $C$ controllable subspace of the system $({\bf A}(t), {\bf B}(t))$, there exist a $\hat C$ controllable subspace of $(\hat {\bf A}, \hat {\bf B})$ such that $\hat C\supset\{\forall \hat v_{(i,t_1)}:v_i\in C\}$. It has been previously shown \cite{HOS80, LIU12a} that a subspace $\hat C$ of a static network is structurally controllable, if there exists a stem-cycle disjoint subgraph that contains all nodes in $\hat C$. A stem is a path starting from a node that is directly coupled to an input signal, in our case these are the intervention points. A stem-cycle disjoint subgraph is a subgraph composed of stems and cycles, such that all nodes are contained by exactly one stem or one cycle. The network representation of the time-independent system is acyclic, hence a stem-cycle disjoint subgraph in our case is simply a set of independent paths. Therefore, $C$ is a structurally controllable subspace of $({\bf A}(t), {\bf B}(t))$, if there exists $\lvert C \rvert$ independent paths starting from intervention points and leading to each node $v_i\in C$ at time $t_1$.
 
\begin{figure}
	\centering
	\includegraphics[scale=.5]{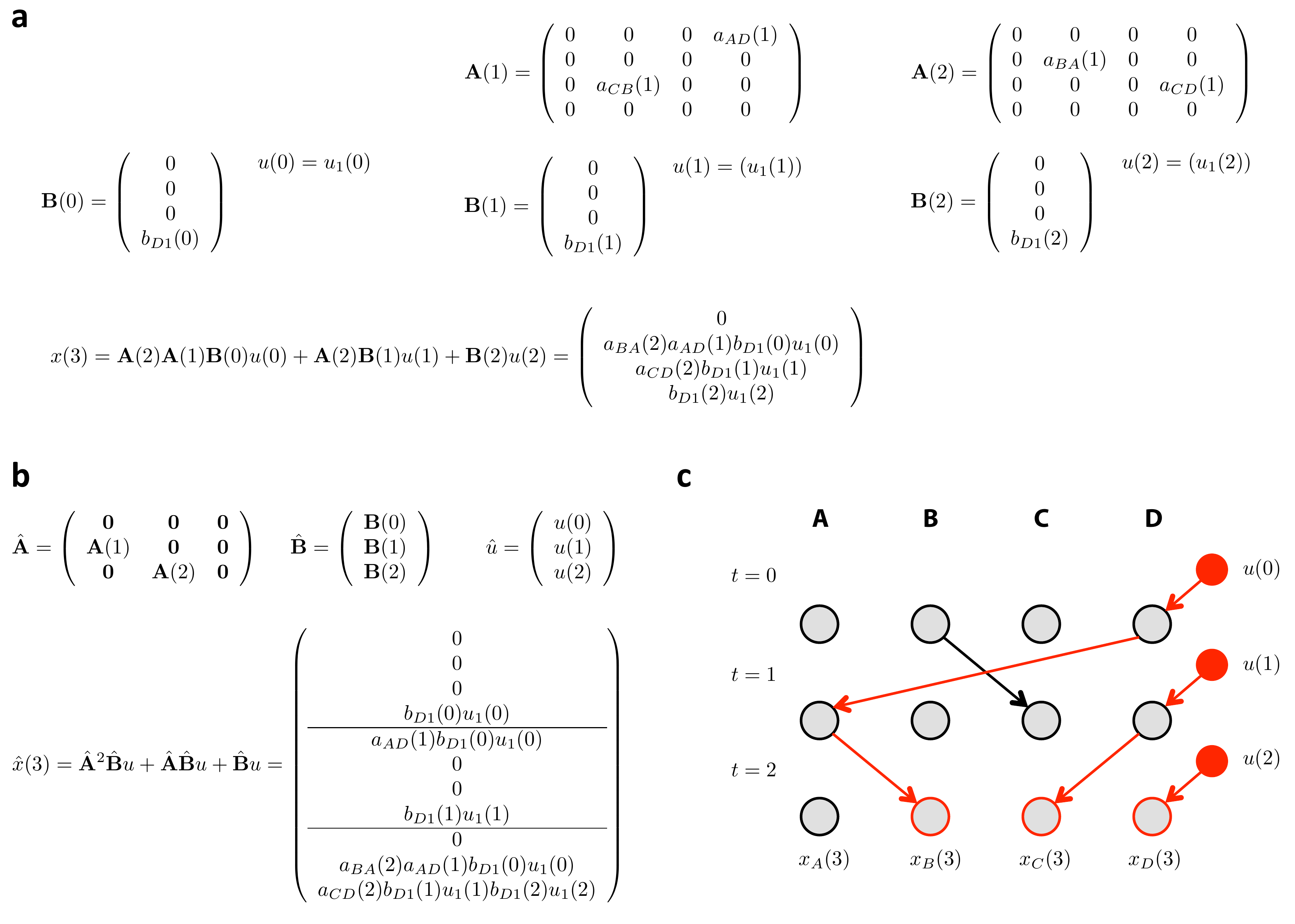}
	\caption{{\bf Converting time dependent dynamics to time invariant.} (a) The matrices describing the linear dynamics corresponding to the example network in Fig.~\ref{fig:layered_def}a. We aim to control the system at target time $t=3$ in $\Delta t=3$ time steps, and we use node $D$ as an input node. We compute $x(3)$ by successfully applying equation~(\ref{eq:model_si}). (b) We convert the linear time-varying dynamics to a larger time-independent system the following way: We create $\hat x \in \mathbb{R}^{N\cdot\Delta t}$ state vector, such that $\hat x_{(i,t)}(\hat t)$ corresponding to $x_i(t)$, $t\in(t_1-\Delta t, t_1]$. We construct $\hat {\bf A}$ by setting $\hat a_{(i,t);(j,t+1)}$ to $a_{ij}(t)$, all other elements of $\hat {\bf A}$ are set to $0$. The input nodes $(D,1)$, $(D,2)$ and $(D,3)$ correspond to the intervention points in the time-varying system. The state vector of the time-independent system $\hat x(t)$ at time $t=3$ is equal to the state vector of the time-varying system $x(t)$. (c) The corresponding network of the time-independent system is equivalent to the layered graph representation $\mathcal L(0, 3)$ of the temporal network $\mathcal T$. According to the independent path theorem, we can control nodes $B$, $C$, and $D$.}
	\label{fig:dynamics_conv}
\end{figure}

 For a small example see Fig.~\ref{fig:dynamics_conv}.

\subsubsection{Maximum controllable subspace problem}

Given a temporal network $\mathcal T$, we select a set of nodes $D\subset V$ to be inputs, meaning that we allow interventions at these nodes. We explore the problem of determining the dimension of the maximum controllable subspace $N_\text C(t_1,\Delta t)$, where $t_1$ is the target time, and $\Delta t$ is the number of time steps we use to reach the desired state. We use the layered representation $\mathcal L(t_1-\Delta t, t_1)$. The set of potential intervention points is $I=\{\hat v_{(i,t)},\forall i,t :v_i\in D\}$ and the set of potential target nodes is $T=\{\hat v_{(i,t_1)},\forall i\}$. A controllable subspace is given by a subset of $T$ for which all nodes can be reached via independent paths from potential intervention points. Therefore, the dimension of the maximum controllable subspace $N_\text C(t_1,\Delta t)$ is the maximum number of independent paths originating from $I$ and terminating in $T$.

The problem of finding the maximum number of independent paths in directed networks is equivalent to solving the maximum flow problem. If the nodes in set $I$ are sources, the nodes in $T$ are sinks, and the capacity of each link and node is set to 1, the maximum flow is equal to the maximum number of independent paths. This problem can be solved in polynomial time, e.g. using the Ford-Fulkerson algorithm with complexity $O(\lvert \hat E\rvert \cdot N_\text C)$~\cite{FOR62, NEW10}.

\subsection{Analytic solution for model networks}

\subsubsection{Temporal network model definition}

We study a simple uncorrelated temporal network model that can be considered as the temporal counterpart of the static hidden parameter model~\cite{CAL02, SOD02}. We start with $N$ unconnected nodes, and for each time step we generate a directed network independently. Each node $v_i$ is assigned two hidden parameters $w_\text{in}(i)$ and $w_\text{out}(i)$. We then randomly place $L$ directed links by choosing the start- and endpoint of the link with probability proportional to $w_\text{in}(i)$ and $w_\text{out}(i)$, respectively. By properly choosing the hidden parameters, we can tune the degree distributions $p_\text{in}(k)$ and $p_\text{out}(k)$. Throughout the paper, we investigate networks with Poisson~\cite{ERD59} and scale-free distribution~\cite{BAR99a}, the latter meaning that the distribution has a power-law tail.

The degree distribution of the model is given by
\begin{equation}
 p(k) = \sum_{i=1}^N \exp\left[{-\frac{w(i)}{\sum_i w(i)}L}\right]\frac{(L w(i)/\sum_i w(i))^k}{k!},
\end{equation}
which is valid for both in- and out-degree. The generating function of this distribution is simply
\begin{equation}
\label{eq:model_generatingfunc}
 G(x)=\sum_{k=0}^\infty p(k)x^k = \sum_{i=1}^N \exp\left[{-\frac{w(i)}{\sum_i w(i)}(1-x)}\right].
\end{equation}

To generate a network with Poisson distribution, we set $w_\text{in}(i)=1$ and $w_\text{out}(i)=1$ for all nodes. This way the probability of connecting any node pair is equal, and we recover the classic Erd{\H o}s-R{\'e}nyi model. The corresponding generating function is
\begin{equation}
 G^\text{ER}(x)= \sum_{i=1}^N \exp\left[{-c(1-x)}\right],
\end{equation}
where $c=L/N=\langle k\rangle$.

To generate networks with scale-free degree distribution we use the so-called static model~\cite{GOH01}. We set the hidden parameters of node $i$ to $w_\text{in/out}(i)=(i+1)^{-\alpha_\text{in/out}}$, where $i=1,2,\ldots,N$. The weights are then shuffled to eliminate any correlations. For large $N$, this choice yields the degree distribution
\begin{equation}
p^{\text{SF}}(k) = \frac{\left[c(1-\alpha)^{1/\alpha}\right]}{\alpha}\frac{\Gamma(k-1/\alpha,c[1-\alpha])}{\Gamma(k+1)}\sim k^{-(1+1/\alpha)} = k^{-\gamma},
\end{equation}
where $c=L/N=\langle k\rangle$ is equal to the average degree, and $\gamma_\text{in/out}=1+1/\alpha_\text{in/out}$ determines the exponent of the tail of the distribution, and $\Gamma(n,x)$ is the upper incomplete gamma function. The corresponding generating function in the $N\rightarrow \infty$ limit is
\begin{equation}
 G^\text{SF}(x)= \frac{1}{\alpha}E_{1+\frac{1}{\alpha}}\left[c(1-\alpha)(1-x)\right],
\end{equation}
where $E_n(x)=\int_1^\infty dt e^{-xt}t^{-n}$ is the exponential integral function. However, for scale-free networks we often find that finite size effects are not negligible for system sizes accessible for simulation. In these cases we have to take the finite size into account by using equation~(\ref{eq:model_generatingfunc}) explicitly.

\subsubsection{Percolation in the temporal network model}

Let us consider the case when we have only one intervention point $\hat v_{(i,t)}$ in the layered network. We can use this intervention to control one of the accessible nodes in a lower layer, i.e. any node that can be reached via a path originating from $\hat v_{(i,t)}$. The cluster of accessible nodes can be described as a cluster generated by the Galton-Watson branching process~\cite{HAR63}: The node $\hat v_{(i,t)}$ has $k_\text{out}$ out-neighbors, where $k_\text{out}$ is drawn from the distribution $p_\text{out}(k)$. Each of these out-neighbors will have $k_\text{out}$ out-neighbors also drawn from  $p_\text{out}(k)$, and so forth.

We study the process in the $N\rightarrow\infty$ limit and denote the probability that the branching process continues forever by $S_\text{out}$. We can calculate $S_\text{out}$ using the self-consistent equation
\begin{equation}
\label{eq:S_out}
 1-S_\text{out} = G_\text{out}(1-S_\text{out}).
\end{equation}
The equation simply means that the probability that the branching process rooted at node $v$ stops in finite steps ($1-S_\text{out}$) is equal to the probability that all branching processes rooted at each out-neighbors of node $v$ also terminte in finite steps. The equation has a trivial solution $S_\text{out}=0$, at the critical point this solution loses stability:
\begin{equation}
 1 = G_\text{out}^\prime(1)=\langle k \rangle,
\end{equation}
meaning that the critical point is simply determined by the average degree independent from other parameters of the degree distribution. Nodes that are roots of infinite trees form the giant out-component.

In the subcritical phase ($\langle k \rangle<1$) the branching process will halt in finite steps, meaning that only finite number of nodes can be accessed. In the critical point ($\langle k \rangle=1$) the size of the largest cluster diverges, however, the relative size is still zero. In the supercritical phase ($\langle k \rangle > 1$) the branching process will continue forever with probability $S_\text{out}$.

Similarly, we calculate the probability that a randomly selected node is an offspring of an infinite cluster:
\begin{equation}
\label{eq:S_in}
 1-S_\text{in} = G_\text{in}(1-S_\text{in}).
\end{equation}
Nodes that are offspring of infinite trees form the giant in-component.

\subsubsection{$N_\text C(\Delta t)$ in the subcritical phase}

The goal of this section is to determine the average $N_\text C(\Delta t)$ using a randomly selected node $v_i$ as input. We start with the observation that if the network is uncorrelated, each intervention point can be treated as independently and randomly selected. In the subcritical phase the size of the accessible cluster rooted at a random node is finite. Therefore, the probability that two such clusters rooted at two randomly selected nodes overlap is 0. The probability that an intervention at $\hat v_{(i,t)}$ can be used to control a node at the target time is equal the probability that a sufficiently long path is rooted at the intervention point. Hence, we first determine the the cumulative distribution function of the maximum path length originating from a randomly selected point, i.e. $P(d)$ is the probability that the maximum length path originating from a node is $\leq d$. The maximum path length from node $\hat v_{(i,t)}$ is 1 larger then the maximum path length originating from its out-neighbors, 
averaging over $p_\text{out}(k)$ we get
\begin{equation}
\label{eq:P(d)}
P(d) = G_\text{out}(P(d-1)).
\end{equation}
We can solve the equation recursively starting from $P(d=0)=p_\text{out}(k=0)$.

Since the model is invariant to time shifts, we can set the control target time to $\Delta t$ without loss of generality. The probability that an intervention at time $t$ can be used is given by  $1-P(\Delta t - t-1)$ for $t\neq \Delta t$, and $1$ is $t=\Delta t$.
Therefore we get
\begin{equation}
\label{eq:N_c(dT)}
N_\text C(\Delta t)= 1 + \sum_{t=1}^{\Delta t-1}1-P(\Delta t-t-1)=1 + \sum_{t=0}^{\Delta t-2} 1-P(t).
\end{equation}
For $\Delta t \rightarrow \infty$ we get:
\begin{equation}
N_\text C = \lim_{\Delta t\rightarrow \infty}N_\text C(\Delta t)= 1 + \langle d \rangle.
\end{equation}

We gain further insight by studying the asymptotic solution of equation~(\ref{eq:P(d)}) for models with Poisson and power-law degree distribution.

For the Poisson case $p_\text{out}(k)$ (or any distribution with finite variance $\sigma_\text{out}^2$), we can expand the generating function $G_\text{out}(x)$ around $x=1$:
\begin{equation}
P(d) = G_\text{out}(1) - G_\text{out}^\prime(1)(1-P(d-1)) = 1 - \langle k \rangle (1-P(d-1)).
\end{equation}
Solving the recursion we get for large $d$
\begin{equation}
\label{eq:P(d)_solution_c<1}
1 - P(d) \sim \langle k \rangle^d,
\end{equation}
where $C$ is some constant. That is $P(d)$ has an exponential tail, e.g. large $d$ values add little to $\langle d\rangle$. This means that $N_\text c(\Delta t)$ approximates its maximum around $\Delta t\sim\langle d \rangle$, and there is little benefit from further increasing $\Delta t$.

However, at the critical point $\langle k \rangle=1$, and thus we need the second-order term in the expansion to extract the asymptotic behavior:
\begin{equation}
\label{eq:P(d)_taylor}
 \begin{aligned}
P(d) &= G_\text{out}(1) - G_\text{out}^\prime(1)(1-P(d-1)) + \frac 1 2 G_\text{out}^{\prime\prime}(1)(1-P(d-1))^2\\
&=  1 - (1-P(d-1)) + \frac{\sigma_\text{out}^2}{2} \left(1-P(d-1)\right)^2.
\end{aligned}
\end{equation}

For large $d$, this yields
\begin{equation}
1-P(d) \sim \frac{2}{\sigma_\text{out}^2}\frac{1}{d}.
\end{equation}
From this it follows that for large $\Delta t$, we get $N_\text c(\Delta t)\sim \log\Delta t$, meaning that increasing $\Delta t$ will increase the number of nodes that we control. However, the fraction of the network that is controlled still remains 0 in the large network limit. 

For scale-free networks with $\gamma<3$, the $\sigma_\text{out}^2$ is infinite, and therefore the simple Taylor series expansion of the generating function in equation~(\ref{eq:P(d)_taylor}) is not sufficient. To understand the effect of a power-law distribution, we transform the generating function provided in equation~(\ref{eq:model_generatingfunc})
\begin{equation}
\begin{aligned}
G_\text{out}(x) &= \frac{1}{\alpha}\left[c(1-\alpha)(1-x)\right]^{1/\alpha}\Gamma\left(-1/\alpha,c[1-\alpha][1-x]\right) \\
&=  \frac{1}{\alpha}\left[c(1-\alpha)(1-x)\right]^{1/\alpha}\left[ \Gamma(-1/\alpha) - \sum_{k=0}^{\infty}\frac{(-1)^k\left[c(1-\alpha)(1-x)\right]^{-1/\alpha+k}}{k!(-1/\alpha+k)} \right].
\end{aligned}
\end{equation}
Using the series expansion form in equation~(\ref{eq:P(d)}) and only keeping the first two terms, we get
\begin{equation}
\begin{aligned}
 P(d+1) &= \frac{1}{\alpha}\left[c(1-\alpha)P(d)\right]^{1/\alpha}\left[ \Gamma(-1/\alpha) + \frac{[c(1-\alpha)P(d)]^{-1/\alpha}}{1/\alpha} +  \frac{ [c(1-\alpha)P(d)]^{-1/\alpha+1}}{1/\alpha-1}\right]\\
  &= 1 + cP(d) + \frac{1}{\alpha}\Gamma(-1/\alpha)\left[c(1-\alpha)P(d)\right]^{1/\alpha}.
\end{aligned}
\end{equation}
Consider $2<\gamma<3$, or equivalently $1/2<\alpha<1$. If $c=\langle k \rangle < 1$, the asymptotic behavior of the solution is determined by the second term, and we obtain the same solution as equation~(\ref{eq:P(d)_solution_c<1}). For the solution in the critical point $\langle k\rangle=1$, we keep the third term, and we find
\begin{equation}
1-P(d) \sim d^{-\alpha/(1-\alpha)} = d^{-1/(\gamma-2)}.
\end{equation}
For $2<\gamma<3$ this means that even in the critical point $N_\text c(\infty)$ will remain finite. However, $N_\text c(\Delta t)$ will approach its stationary value slowly, that is $N_\text c(\infty)-N_\text c(\Delta t)\sim \Delta t^{-(3-\gamma)/(\gamma-2)}$.

\subsubsection{$n_\text C(\Delta\tau)$ in the supercritical phase}

Above the critical point, the probability that an intervention point is a root of an infinite tree is $S_\text{out}>0$, meaning that there exists infinite length paths originating from the node. As a consequence, by choosing $\Delta t= \Delta\tau N$ ($\Delta\tau>0$) we can control finite fraction of the network $n_\text c(\Delta\tau)=N_\text c(\Delta\tau N) / N$ using infinite paths, and the contribution of finite size clusters is negligible.

Consider the case when $v_i$ is the input node, and $v_{(i,t)}$ and $v_{(i,t^\prime)}$ are two intervention points such that both are roots of infinite trees. Since these trees cover finite fraction of the network, we can no longer assume that the overlap of accessible clusters has zero probability. However, being the root of an infinite tree also means that we can reach a finite fraction $S_\text{out}$ of the nodes in the target layer, and we can choose from many possible paths. Therefore, for small $\Delta\tau$, we assume that whenever an intervention point is a root of an infinite tree, we can use that intervention point to control one node in the target layer. This means that
\begin{equation}
\label{eq:nc_low_tau}
 n_\text c(\Delta\tau) = S_\text{out}\Delta\tau .
\end{equation}
Note that this does not depend on the in-degree distribution $p_\text{in}(k)$.

\begin{figure}
	\centering
	\includegraphics[scale=.5]{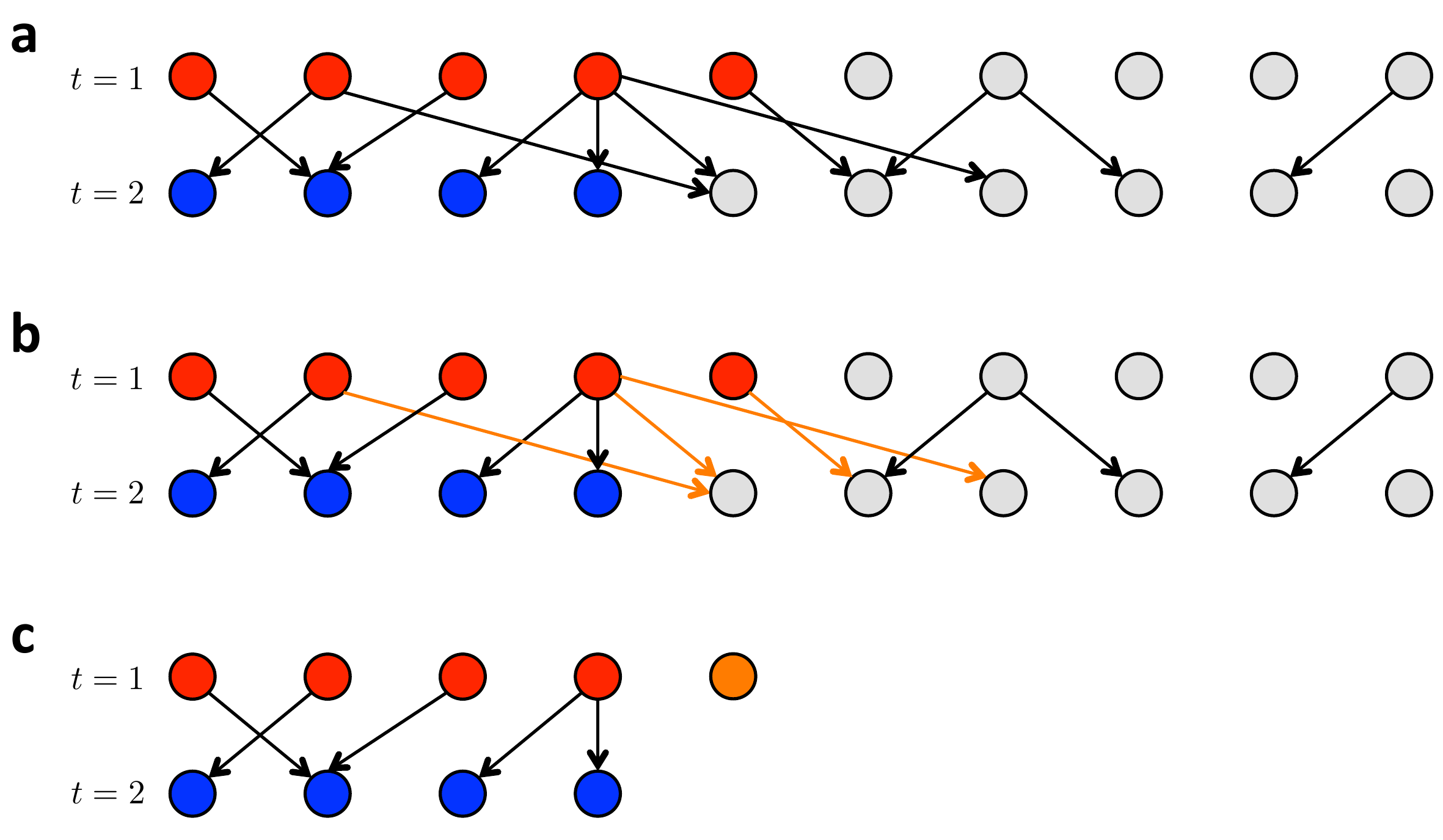}
	\caption{{\bf Determining the degree distribution within the giant component.} (a) The goal is to determine the degree distribution within the giant component, i.e. nodes that belong to both the giant in-component and the giant out-component. We consider two consecutive layers $t=1$ and $t=2$. The red nodes are the giant in-component in layer $t=1$, and the blue nodes are the giant out-component in layer $t=2$. The giant component in layer $t=1$ consists of the red nodes that are connected to at least one blue node. (b) First, we remove all links that lead to nodes outside the giant component, i.e. the grey nodes. The network is uncorrelated, therefore the link link removal can be treated as random. (c) Next, we remove the now isolated red nodes, i.e. nodes that are in the in-component, but not in the out-component.}
	\label{fig:giantcompdegdist}
\end{figure}

For large $\Delta\tau$, $n_\text c(\Delta\tau)$ saturates, since it is limited by the maximum throughput of the giant component, i.e. the maximum number of independent infinite paths. The giant component consists of nodes in each layer that are both in the giant in-component and the giant out-component. To calculate $n_\text c(\Delta\tau)$ the first step is to determine the degree distributions $\tilde p_\text{in/out}(k)$ within the giant component. Consider two adjacent layer of nodes at $t=1$ and $t=2$ (Fig.~\ref{fig:giantcompdegdist}), the two layers are connected by links that are active at time $t=1$. We aim to determine $\tilde p_\text{out}(k)$ for $t=1$. The nodes that are in the giant component in layer $t=1$ are the nodes that are in the giant in-component, and have at least one connection to nodes in layer $t=2$ that are in the giant out-component. First, we remove the links connecting the in-component with nodes not in the out-component, this is equivalent to randomly removing $1-S_\text{in}$ links. Now all nodes that have at 
least one connection left are members of the giant component. Therefore, to obtain $\tilde p_\text{out}(k)$ we remove the nodes with $0$ connections. This leads to
\begin{equation}
\label{eq:randomremoval}
\tilde p_\text{out}(k) = (1-\delta_{k,0})\sum_{j=k}^{\infty} {j\choose k} S_\text{in}^k (1-S_\text{in})^{j-k},
\end{equation}  
and we will use the corresponding generating function
\begin{equation}
\tilde G_\text{out}(k) = \frac{G_\text{out}(1-S_\text{in} + S_\text{in}x) - G_\text{out} (1-S_\text{in})}{1-G_\text{out} (1-S_\text{in})}.
\end{equation}
The in-degree distribution is determined similarly.

\begin{figure}
	\centering
	\includegraphics[scale=.5]{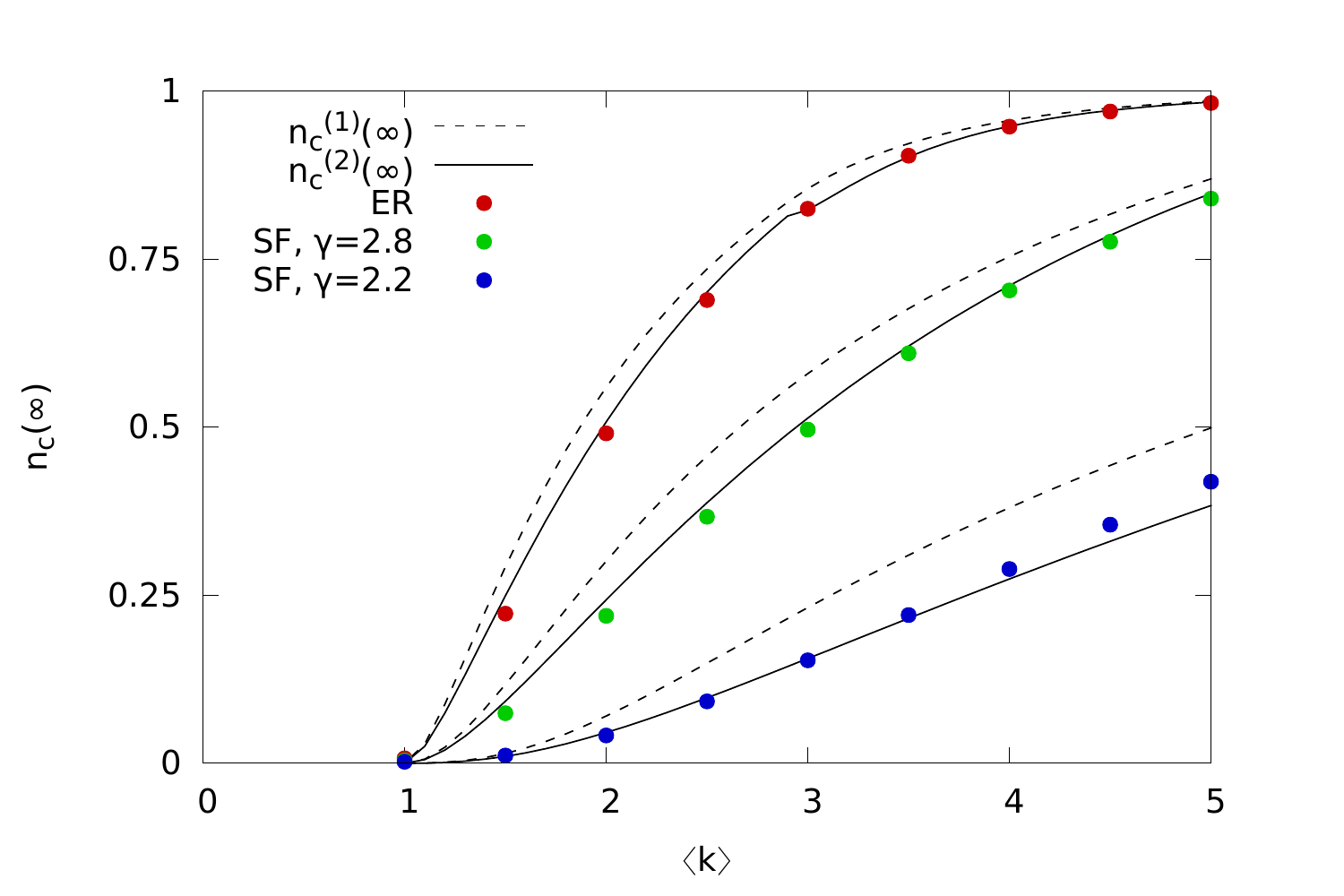}
	\caption{{\bf Comparing the approximations.} We plot the average controllable fraction of the network in function of the average degree for Poisson and scale-free degree distributions. for networks of size $N=1,000$, each data point is an average of $1,000$ input node measurements. The dashed line is the first approximation $n^{(1)}_\text C(\infty)$, and the solid line is the second analytical approximation $n^{(2)}_\text C(\infty)$. The small break in $n^{(2)}_\text C(\infty)$ for the Erd\H os-R\'enyi network is a consequence of the core percolation transition~\cite{LIU12}, the transition point for scale-free networks is not shown on the plot.}
	\label{fig:compareapprox}
\end{figure}

\begin{figure}
	\centering
	\includegraphics[scale=.5]{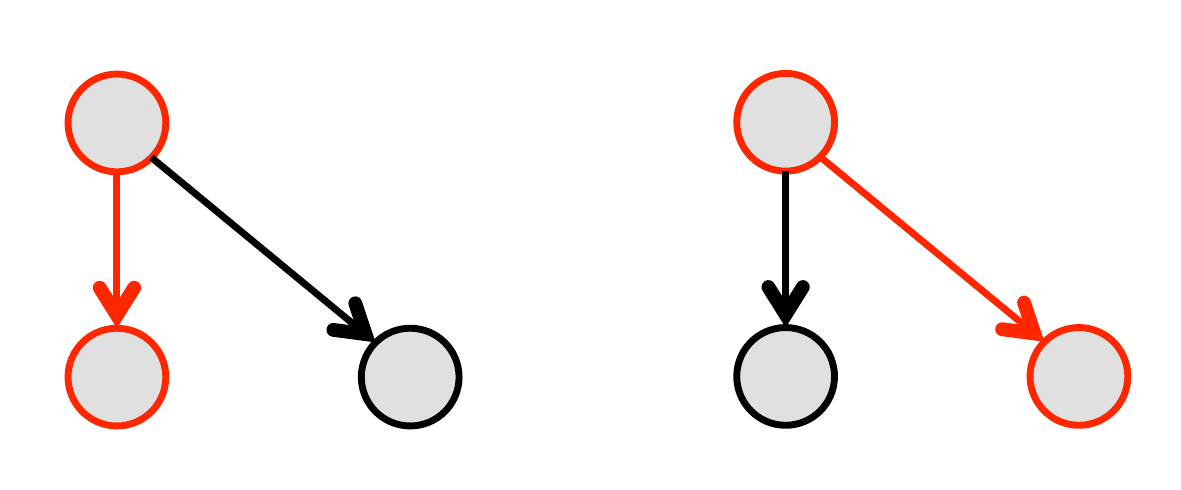}
	\caption{{\bf Role of nodes in possible maximum matchings.} We show a small example network of 3 nodes. In this network the size of the maximum matching is one, and there are two possible configurations highlighted in red. The top node is always matched, and the two bottom nodes cannot be matched in the same configuration.}
	\label{fig:nodecateg}
\end{figure}

To calculate a first approximation  $n^{(1)}_\text C(\infty)$, we determine the maximum number of independent paths in the giant component connecting two subsequent layers $t=1$ and $t=2$, which is equivalent to finding the maximum matching in a bipartite network formed by the two layers. A matching in a network is defined as a set of links that do not share endpoints, therefore in the case of the network of two layers, the links in the matching are independent paths of length one. A node is called matched, if they are adjacent to a link in the matching. This way $n^{(1)}_\text C(\infty)$ is equal to the maximum matching in a bipartite network with $S_\textbf{in}S_\textbf{out}N$ nodes in each layer, and degree distributions $\tilde p_\text{out}(k)$ and $\tilde p_\text{in}(k)$. For uncorrelated networks the size of the maximum matching can be determined analytically, we provide the detailed calculation in Sec.~\ref{sec:matching}. This approximation yields an upper bound for $n_\text C(\infty)$ (Fig.~\ref{fig:compareapprox}), 
because it assumes that we can choose the endpoints of the paths in layer $t=1$, and the starting points of the paths in layer $t=2$ arbitrarily. However, when constructing a maximum matching we do not have such freedom: some nodes always have to be matched~\cite{JIA13}, and in other cases some nodes cannot be included at the same time, for a small example see Fig.~\ref{fig:nodecateg}.

\begin{figure}
	\centering
	\includegraphics[scale=.5]{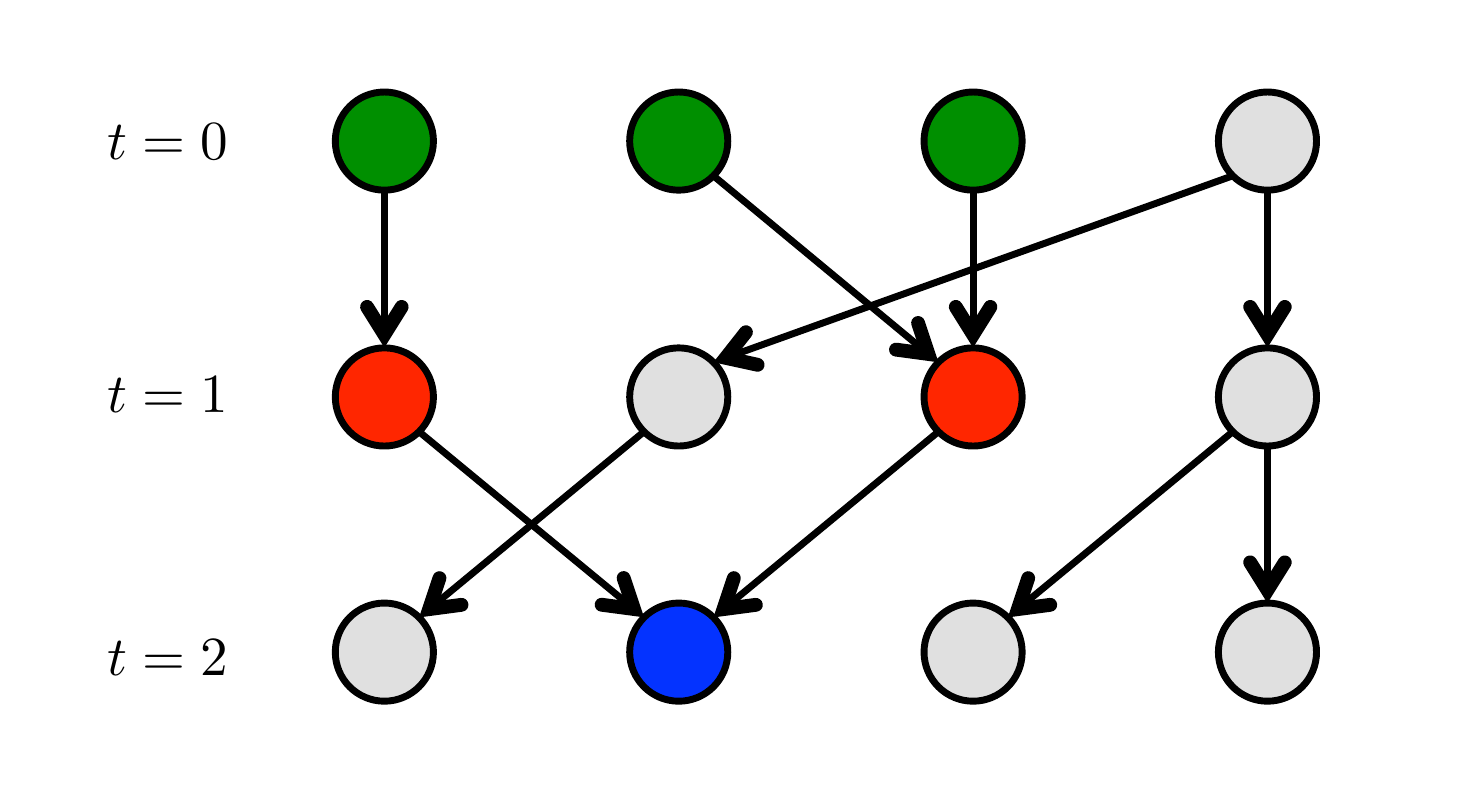}
	\caption{{\bf Second approximation of }$n_\text C(\infty)${\bf .} Consider the giant component in three consecutive layers $t=0,1,2$. Examining the maximum matching in layers $t=0$ and $t=1$, we find that the red nodes are matched in all possible maximum matchings. However, in the next layer the red nodes are connected to the same node, hence they cannot be matched at the same time. Therefore, the two red nodes can only be used in one independent path. Counting such configurations provides the second approximation.}
	\label{fig:approx2}
\end{figure}

For the next approximation $n^{(2)}_\text C(\infty)$, we consider three subsequent layers $t=0,1,2$. Each layer contains $S_\textbf{in}S_\textbf{out}N$ nodes, and has degree distributions $\tilde p_\text{out}(k)$ and $\tilde p_\text{in}(k)$. First, we examine the maximum matching between layers $t=0$ and $t=1$, and we determine the set of nodes $A$ in layer $t=1$ that are  matched in all possible the maximum matchings. In the first approximation, these nodes will always be endpoints of independent paths. However, if we cannot match them in the next layer, they will become dead ends. Therefore, the number of nodes in $A$ that cannot be matched at the same time will be the next correction to $n^{(1)}_\text C(\infty)$~(Fig.~\ref{fig:approx2}). To calculate the correction, we find the maximum matching in the bipartite network formed by nodes in $A$ in layer $t=0$, and all nodes in layer $t=1$. The degree distribution of nodes in $A$ is $\tilde p_\text{out}(k)$, and the degree distribution of nodes in layer $t=2$ can be calculated by randomly 
removing $1-\lvert A \rvert /S_\textbf{in}S_\textbf{out}N$ fraction of links from $p_\text{in}(k)$, similarly to equation~(\ref{eq:randomremoval}). The number of nodes in $A$ is determined using the equations developed in~\cite{JIA13}.

Similar correction can be computed for the set of nodes $B$ in layer $t=1$ that are always matched from layer $t=2$, but cannot be matched at the same time from layer $t=1$. We find that $N^{(2)}_\text C(\infty)$ approximates the numerical simulations well (Fig.~\ref{fig:compareapprox}).

{\it Note:} In~\cite{JIA13}, it was shown that for dense networks above the core percolation threshold, the number of nodes that are always matched can be drastically different depending on specific realization of the network model, e.g. two Erd\H os-R\'enyi networks generated with the same parameters can be different. This is due to a special case, when a finite fraction of nodes are "almost always" matched, meaning that we have a set of nodes $A$ such that in each possible matching only a finite number of nodes in $A$ are not matched. Therefore, for our purposes these nodes can be treated as always matched.

\subsubsection{Matching in bipartite networks}\label{sec:matching}

In this section we calculate the relative size of the maximum matching in uncorrelated bipartite networks with arbitrary degree distribution. Let $\mathcal B$ be a bipartite network, with two sets of nodes $V^-$ (lower) and $V^+$ (upper) and a set of links $E$, such that there each link connects one upper node $v^+\in V^+$ and one lower node $v^-\in V^-$. $p^+(k)$ and $p^-(k)$ are the degree distributions of the upper and lower sides, respectively. We use the notations $N^+=\lvert V^+ \rvert$, $N^-=\lvert V^- \rvert$ and $L=\lvert E\rvert$. The average degree of each layer is $c^\pm=L/N^\pm$. If $M$ is the set of links in the matching, we define $m^\pm=\lvert M \rvert / N^\pm$. The maximum matching problem for bipartite networks have been studied for the case when $N^+=N^-$~\cite{LIU11}, here we extend the solution to the $N^+\neq N^-$ case.

We use the formalism developed for core percolation~\cite{LIU12}. Core percolation describes the sudden emergence of the core in random networks~\cite{KAR81, BAU01, ZDE06}. To define the core, we first introduce the greedy leaf removal (GLR) process: we select a leaf randomly (a node with degree 1), and remove that node and its neighbor together with all links adjacent to that neighbor, we repeat this step until no leaves are left; we then remove all isolated nodes. The core is defined as the remainder of the network after the GLR.

Analytic description is possible by introducing the following node categories: (i) $\alpha$-removable, nodes that can become isolated during the GLR; (ii) $\beta$-removable, nodes that can be removed as a neighbor of a leaf during the GLR. We define $\alpha^+$ as the probability that following a random link to the upper side we find a node that is $\alpha$-removable in the absence of the link. We define  $\alpha^-$, $\beta^+$, and $\beta^-$ similarly. These probabilities are determined by a set of self-consistent equations:
\begin{align}
 \alpha^\pm&=H^\pm(\beta^\mp), \\
 \beta^\pm&=1-H^\pm(1-\alpha^\mp),
\end{align}
where $H^\pm(x)=\sum_{k=1}^\infty k p^\pm(k)/\langle k\rangle x^{k-1}$ is the generating function of the excessive degree distribution.

The GLR process can be used to construct a maximum matching in the class of bipartite networks that we study here. We remove a leaf that consist of node $v_1$ with degree 1, and node $v_2$ with possibly higher degree.  To construct the maximum matching, we add link $(v_1, v_2)$ to the matching. Now all links adjacent to $v_2$ are not allowed in the matching, and therefore, we remove them too. We can continue, until we have no leaves left, i.e. we are left with the core. It was shown that in large non-bipartite random networks, the core can be asymptotically matched, i.e. the probability of randomly choosing an unmatched node is 0~\cite{ZDE06}. However, in bipartite networks there is another limiting factor: if the size of the core is different on the two sides, the size of the matching in the core cannot be larger then the smaller side.

Note that this can also happen if $N^+=N^-$, but $p^+(k)\neq p^-(k)$), and this limitation was not considered in~\cite{LIU11} and in the subsequent~\cite{POS13}. Therefore, their results should be cautiously applied to networks above the core percolation with asymmetric degree distributions.

As stated above, $m^\pm$ is the sum of the contribution of the leaf removal and the core. We first calculate the contribution of leaf removal. For each leaf removal, we add one link to the matching, increasing the number of matched nodes by 2, one on both sides. For each $\beta$-node there is one leaf removal. Therefore, to calculate the contribution of the leaf removal, we count the $\beta$-nodes on both sides:
\begin{equation}
 N_\beta^\pm=N^\pm [1-G^\pm(1-\alpha^\mp)].
\end{equation}
However, by doing this we have double counted the case when two $\beta$-nodes are removed together. This can only happen, if in the absence of the link connecting the two nodes, both nodes are $\alpha$-nodes, the probability of this event is $\alpha^+\alpha^-$ for each link. Therefore, the overall contribution is
\begin{equation}
 N^+ [1-G^+(1-\alpha^-)] + N^- [1-G^-(1-\alpha^+)] - L \alpha^+\alpha^-.
\end{equation}

To determine the contribution of the core, we calculate the size of the core on both sides:
\begin{equation}
 N_\text{core}^\pm= N^\pm\left[ G^\pm(1-\alpha^\mp) -G^\pm(\beta^\mp) - c^\pm\alpha^\pm(1-\beta^\mp-\alpha^\mp) \right],
\end{equation}
and select the smaller side. Therefore, all together we have
\begin{subequations}
\begin{align}
 m^-&= \frac{1}{N^-} \left(N^- [1-G^-(1-\alpha^+)] + N^+ [1-G^+(1-\alpha^-)] - L \alpha^+\alpha^- + \min_\pm N_\text{core}^\pm\right)\\
 &= [1-G^-(1-\alpha^+)] + \frac{N^+}{N^-}[1-G^+(1-\alpha^-)]- c^- \alpha^+\alpha^- +  \frac{1}{N^-}\min_\pm N_\text{core}^\pm.
\end{align}
\end{subequations}

\subsection{Dataset analyzed}

\subsubsection{Description}

We study a publicly available temporal network representing the email communication of a mid-size company~\cite{MIC11, konect}. The data set contains the sender, the recipient, and the time each email has been sent. All together there are 82,927 emails between 167 employees covering a 9 month period.

The necessary temporal resolution of the network depends on the time scale of the dynamical process we aim to control. To highlight different features of the dataset, we use two different temporal resolutions with one hour and one day time steps. To obtain these networks, we preform a coarse graining procedure: for each time step $t_0\leq t < t_1$ we create an aggregated network, i.e. we connect nodes $v_i$ and $v_j$ in the coarse grained network, if at least one email has been sent between $t_0$ and $t_1$.

The one hour coarse grained network corresponds to a scenario, when we aim to influence the dynamics within a day. An important feature of the data set is that it follows strong daily and weekly patterns. The bulk of the email traffic happens during the 9 hour period of the regular office hours on workdays (Fig.~\ref{fig:real_hour}a-b). The average degree of the network outside the working hours is approximately 0, while during the office hours $\langle k\rangle_\text h \approx 0.23$. This means that control on the hourly time scale is only possible within one day, that is each day can be considered separately. We find that the average degree distribution is highly heterogeneous (Fig.~\ref{fig:real_hour}c-d), the second moment ($\langle k_\text{out} ^2\rangle_\text{h}\approx 0.99$, $\langle k_\text{in} ^2\rangle_\text{h}\approx 0.36$) is much larger than the second moment of a Poisson distribution with the same average degree ($\langle k ^2\rangle_\text{ER}\approx 0.28$).

By choosing one day time steps we assume slower dynamics on the network. The coarse graining removes the daily activity patterns. To study control spanning over multiple weeks, we explicitly remove weekends and holidays, i.e. we measure the time in workdays. The average degree within a time step is $\langle k\rangle_\text D \approx 1.76$ (Fig.~\ref{fig:real_day}a), which predicts that the system is in the supercritical phase, meaning that the characteristic control time is in the order of the system size. Therefore, the length of the available time period does not allow multiple independent measurements of the control process, hence we focus on controlling the system at the end of the last workday at $t=190$. Similarly to the one hour case, the average degree distribution within a time step is heterogeneous (Fig.~\ref{fig:real_day}b-c), with second moments $\langle k_\text{out} ^2\rangle_\text{h}\approx 18.07$ and  $\langle k_\text{in} ^2\rangle_\text{h}\approx 11.12$) compared to the second moment assuming a Poisson distribution with the same average degree $\langle k ^2\rangle_\text{ER}\approx 4.86$.

\subsubsection{Randomization procedures}

We use four different randomization techniques to identify which temporal or network characteristics of the system influence controllability.

\begin{figure}
	\centering
	\includegraphics[scale=.45]{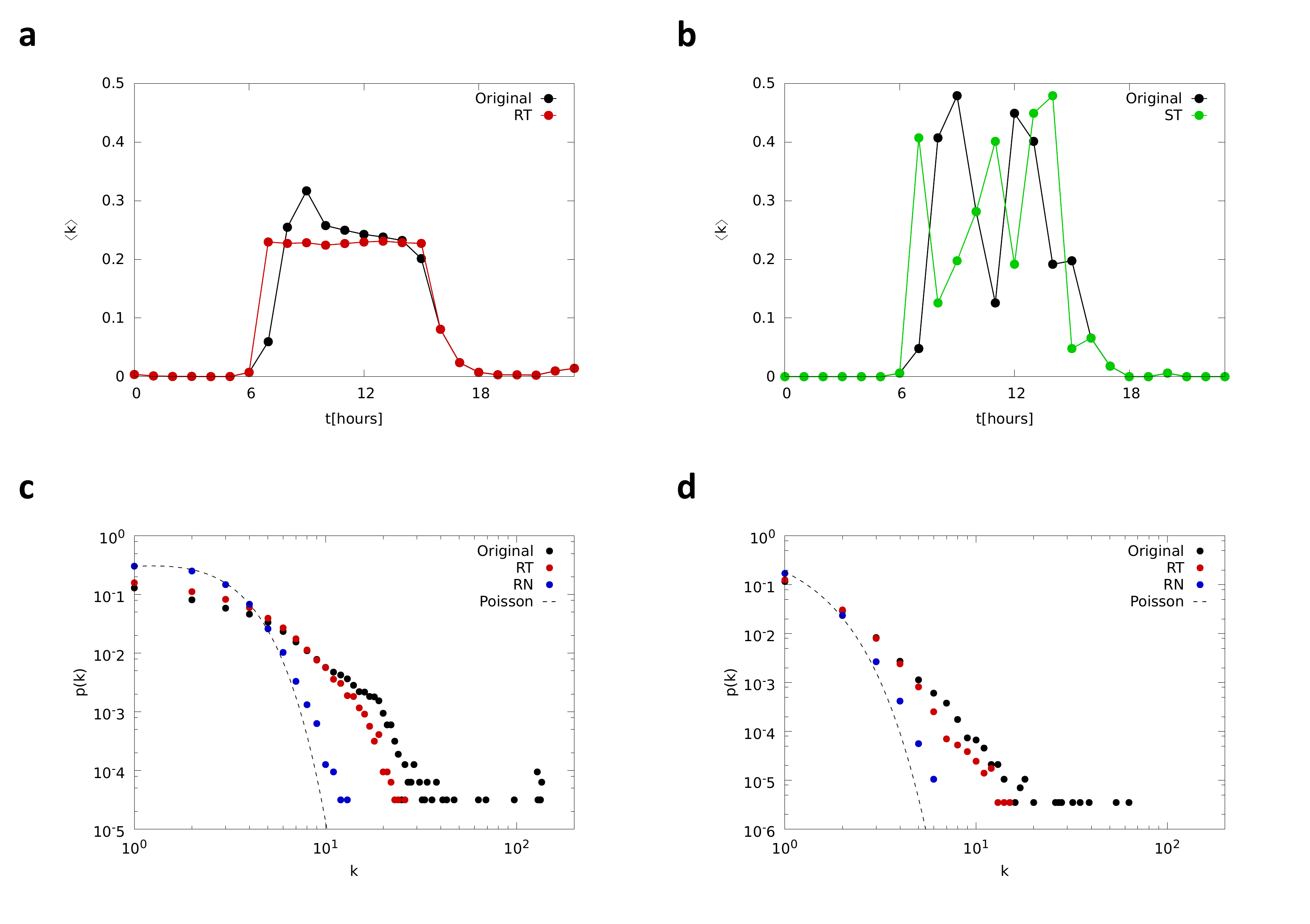}
	\caption{{\bf The temporal network of email communication with one hour time resolution.} (a) The average degree of the network at different hours of a day, the plot shows the average of all workdays. The bulk of the email traffic happens during a 9 hour period corresponding to the office hours (black). We eliminate the fluctuations in the average degree by assigning random times to the links within the 9 hour active period (RT, red). (b) The average degree of the network at different hours of a typical workday. The average degree is characterized by large fluctuations. Shuffling the time steps keeps the fluctuations, but eliminates correlations between consecutive layers (ST, green). (c-d) The in- and out-degree distribution within a time step, the plot shows the average of all working hours. The original distributions are highly heterogeneous (black) compared to the Poisson distribution with the same average. Assigning random link times indirectly changes the distributions, however, the distributions remain heterogeneous (RT, red). Randomizing the network topology within a time step eliminates the heterogeneity (RN, blue). Note that the RN randomization keeps the fluctuations in the average degree, hence the difference from the Poisson distribution. To obtain the Poisson distribution, we have to apply both the RT and the RN randomization.}
	\label{fig:real_hour}
\end{figure}

\begin{figure}
	\centering
	\includegraphics[scale=.45]{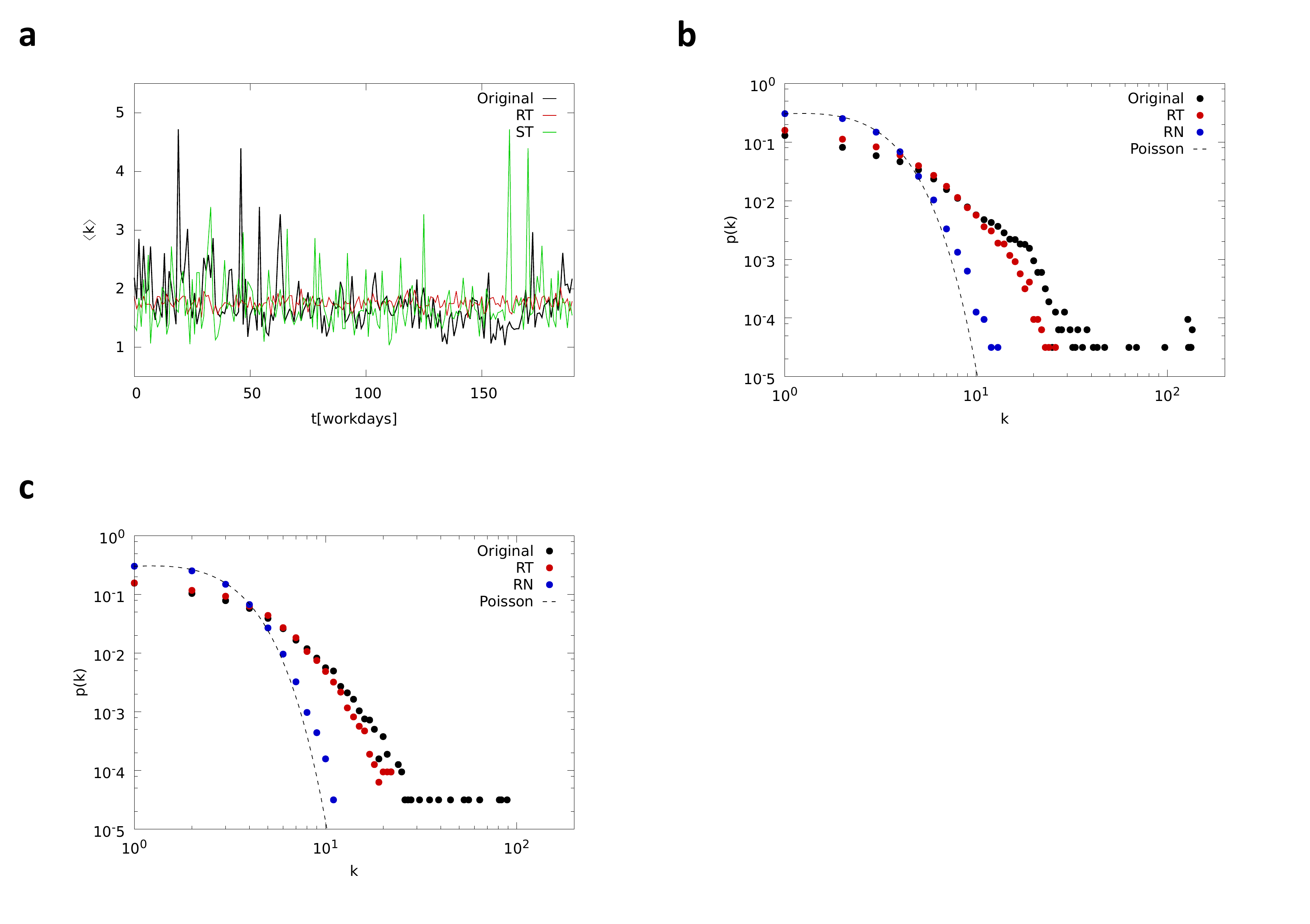}
	\caption{{\bf The temporal network of email communication with one day time resolution.} (a) The average degree of the network at different days. Similarly to the one hour case, the average degree has large fluctuations. Randomly assigning link times eliminates the fluctuations (RT, red), and shuffling the time steps only eliminates correlations between consecutive time steps (ST, green). (b-c) The in- and out-degree distribution within a time step, the plot shows the average of all working hours. The original distributions are highly heterogeneous (black) compared to the Poisson distribution with the same average. Assigning random link times indirectly changes the distributions, however, the distributions remain heterogeneous (RT, red). Randomizing the network topology within a time step eliminates the heterogeneity (RN, blue).}
	\label{fig:real_day}
\end{figure}

\emph{Random time (RT)}: This randomization assigns random time steps to each link, thereby removing all temporal correlations, both overall fluctuations in the average degree, and local correlations such as consequent and simultaneous events (Fig.~\ref{fig:real_hour}a and \ref{fig:real_day}a). This randomization does not change who interacts with whom, that is it does not change the aggregated network. However, by separating simultaneous events, the randomization changes the degree distribution within a time step indirectly (Fig.~\ref{fig:real_hour}c-d and \ref{fig:real_day}b-c). For the one hour coarse grained network, we only randomize within the working hours of each workday.

\emph{Shuffled time (ST)}: We shuffle the time steps, removing all correlations between subsequent time steps, such as casual chain of events, the structure within a time steps remains unchanged (Fig.~\ref{fig:real_hour}b and \ref{fig:real_day}a). For the one hour coarse grained network, we only shuffle the time steps within the working hours of each workday.

\emph{Random network (RN)}: In this randomization, the network for each time step is replaced by an Erd{\H o}s-R\'enyi network with the same number of links, thereby removing all network structure, including the heterogeneity from the degree distribution (Fig.~\ref{fig:real_hour}c-d and \ref{fig:real_day}b-c). All interaction times are retained, preserving the fluctuations in the average degree.

\emph{Degree preserved network (DPN)}: For this randomization, we break all connections, and randomly rewire them within a time step. This way only the degree distribution is preserved, but all other correlations in the network structure are eliminated. Similarly to RN, we do not change the interaction times.

\newpage

\end{document}